\newif\ifsubmode
\submodefalse
\ifsubmode
  \documentstyle[12pt,aasms4,epsf]{article}
  \received{}
  \accepted{}
  \journalid{}{}
  \articleid{}{}
\else
  \documentstyle[11pt,aaspp4,epsf]{article}
  \slugcomment{{\sl accepted for publication in The Astrophysical Journal}}
\fi

\lefthead{Cretton et al.}
\righthead{The Distribution of Stellar Orbits in the Giant Elliptical Galaxy 
NGC 2320}

\def\etal{{et al.~}}
\def\eg{e.g.,~}
\def\ie{i.e.,~}

\def\Lsun{{\rm\,L_\odot}}
\def\kms{{\rm\,km\,s^{-1}}}

\def\xp{x'}
\def\yp{y'}

\def\Reff{\rm R_{eff}}

\def\vlos{v_{\rm los}}

\def\gas{{\rm gas}}    
\def\sin{{\rm sin}}  
\def\exp{{\rm exp}}
\def\deg{^{\rm o}}
 
\def\spose#1{\hbox to 0pt{#1\hss}}
\def\lta{\mathrel{\spose{\lower 3pt\hbox{$\sim$}}
    \raise 2.0pt\hbox{$<$}}}
\def\gta{\mathrel{\spose{\lower 3pt\hbox{$\sim$}}
    \raise 2.0pt\hbox{$>$}}}

\begin{document}

\title{The Distribution of Stellar Orbits in the Giant Elliptical Galaxy 
NGC 2320\footnote{Observations reported in this paper were obtained at 
the Multiple Mirror Telescope Observatory, a facility operated jointly by 
the University of Arizona and the Smithsonian Institution, and at the
KPNO 4 meter telescope which is operated by AURA, Inc. under cooperative
agreement with the National Science Foundation.}}

\author{Nicolas Cretton}
\affil{Sterrewacht Leiden, Postbus 9513, 2300 RA Leiden, The Netherlands}
\affil{Max-Planck-Institut f\"ur Astronomie, K\"onigstuhl 17, 69117 
Heidelberg, Germany\altaffilmark{2}}

\author{Hans--Walter Rix}
\affil{Max-Planck-Institut f\"ur Astronomie, K\"onigstuhl 17, 69117 
Heidelberg, Germany}

\author{P.~Tim de Zeeuw,}
\affil{Sterrewacht Leiden, Postbus 9513, 2300 RA Leiden, The Netherlands}

\altaffiltext{2}{present address}

%%%%%%%%%%%%%%%
% Start the abstract on a fresh page
%%%%%%%%%%%%%%%

\ifsubmode\else
\clearpage\fi

%%%%%%%%%%%%%%%
% Use a small baselineskip, unless in submission mode.
%%%%%%%%%%%%%%%

\ifsubmode\else
\baselineskip=14pt
\fi

%%%%%%%%%%%%%%%
% Abstract
%%%%%%%%%%%%%%%

\begin{abstract}
 
We present direct observational constraints on the orbital
distribution of the stars in the giant elliptical NGC 2320. Long-slit
spectra along multiple position angles are used to derive the stellar
line--of--sight velocity distribution within one effective radius. In
addition, the rotation curve and dispersion profile of an ionized gas
disk are measured from the [OIII] emission lines.  After correcting
for the asymmetric drift, we derive the circular velocity of the gas,
which provides an independent constraint on the gravitational
potential.

To interpret the stellar motions, we build axisymmetric
three--integral dynamical models based on an extension of the
Schwarzschild orbit--superposition technique.  We consider two
families of gravitational potential, one in which the mass follows
the light (\ie no dark matter) and one with a logarithmic
gravitational potential.

Using $\chi^2$--statistics, we compare our models to both the stellar
and gas data to constrain the value of the V-band mass--to--light
ratio $\Upsilon_V$.  We find $\Upsilon_V = 15.0\pm 0.6$~$h_{75}$ for
the mass-follows-light models and $\Upsilon_V = 17.0\pm 0.7$~$h_{75}$
for the logarithmic models.  For the latter, $\Upsilon_V$ is defined
within a sphere of $15''$ radius.

Models with radially constant $\Upsilon_V$ and logarithmic models with
dark matter provide comparably good fits to the data and possess
similar dynamical structure. Across the full range of $\Upsilon_V$
permitted by the observational constraints, the models are radially
anisotropic in the equatorial plane over the radial range of our
kinematical data ($1''
\lta r \lta 40''$). Along the true minor axis, they are more nearly
isotropic. The best fitting model has $\sigma_r/\sigma_{\rm total}
\simeq 0.7$, $\sigma_\phi/\sigma_{\rm total} \simeq 0.5-0.6$ and
$\sigma_\theta/\sigma_{\rm total} \simeq 0.5$ in the equatorial plane.
\end{abstract}

\section{Introduction}

The orbital distribution of stars, often quantified by the
(an)--isotropy of the local velocity dispersion, is a useful measure
of the dynamical state of a galaxy.  It provides constraints on galaxy
formation and helps discriminating between N-body merger
remnants. Furthermore, it has been a classic diagnostic to study the
mass distribution of galaxies.

Anisotropy profiles can only be obtained from observations through a
dynamical model, since only line--of--sight velocities can be measured
for external galaxies. For a number of reasons the construction of
such models is more difficult for elliptical galaxies than for
spirals.  Elliptical galaxies generally have no simple tracer
population that could be used to derive the underlying gravitational
potential (like HI disks in spirals). These systems are known to
support a variety of orbital shapes that largely overlap each
other. Furthermore different orbit distributions and gravitational
potentials may result in very similar observable kinematics, known as
the mass--anisotropy degeneracy.  On the observational side, stellar
velocity distributions in ellipticals usually have to be derived from
absorption lines, which limits the maximal radius for kinematics.
 
This situation is improving on various fronts: fully anisotropic
dynamical models can now be constructed and matched to observations
(see \eg Rix \etal 1997; Richstone \etal 1997; Gerhard \etal 1998;
Cretton \etal 1999, hereafter C99; Gebhardt \etal 2000). The
mass--anisotropy degeneracy can in principle be broken with the use of
the full line--of--sight velocity distributions, also called Velocity
Profiles (hereafter VPs) (see \eg Dejonghe 1987; Gerhard
1993). Measurements of VPs at several position angles and extended
radii further constrain the dynamical structure of the galaxy (see \eg
Carollo \etal 1995; Statler, Smecker--Hane \& Cecil 1996).

As a consequence, information about the intrinsic dispersion profiles
in elliptical galaxies is slowly emerging: spherical anisotropic
models for NGC 2434 show $\sigma_r/\sigma_{\rm total} = 0.7$,
$\sigma_\theta/\sigma_{\rm total} = \sigma_\phi/\sigma_{\rm total} =
0.5$.  Matthias \& Gerhard (1999) built axisymmetric 3--integral
models of NGC 1600 (they did not need any DM inside 1 effective radius
$\Reff$, enclosing half of the light), deducing radial anisotropy in
the outer parts ($\sigma_\theta/\sigma_r
\simeq \sigma_\phi/\sigma_r \simeq 0.7$), and a more isotropic
structure in the center.  In their spherical models of NGC 6703,
Gerhard \etal (1998) again found near isotropy in the center, and a
mild radial anisotropy in the outer parts, as in NGC 1399 (Saglia
\etal 1999). Dejonghe \etal (1996) inferred tangential anisotropy in
NGC 4697, as in NGC 1700 (Statler \etal 1999) based on axisymmetric
3--integral models built with a quadratic programming technique, but
no VPs were used in these two studies.  Using fully general
axisymmetric models, van der Marel \etal (1998) explored the orbital
structure of the small rapidly rotating elliptical M32; using similar
technique, Gebhardt \etal (2000; and in preparation) studied NGC 3379, NGC
3377, NGC 4473 and NGC 5845 and found that they were radially
anisotropic in the range (0.1 - 1) $\Reff$.  Merritt \& Oh (1987)
derived a slight radial anisotropy $\sigma_r \simeq 1.2
\sigma_{\phi} = 1.2 \sigma_{\theta}$ in the main body of M87, but they
did not model the full VPs.

In this paper, we derive the anisotropy of the stellar orbits in the
giant elliptical NGC 2320, combining constraints from the stellar VPs
and from the gas rotation and dispersion curves. The simple kinematics
of the gas (mainly circular rotation) allows an independent and
straightforward measure of the total mass of the system. Cinzano \&
van der Marel (1994) constructed dynamical models for NGC 2974 and
found that the velocities of the ionized gas disk were consistent with
the potential derived from the stellar kinematics.

The paper is organized as follows. In Section~\ref{s:datatwentytwo},
we present the photometric and kinematic data. We describe the mass
and dynamical models in Section~\ref{s:models}. The range of
statistically acceptable models with their associate dynamical
structure is derived in Section~\ref{s:resultstwentytwo}.
Section~\ref{s:conclusionstwentytwo} summarizes the results and
discusses them in the context of cosmological simulations and mergers.

\section{The data\label{s:datatwentytwo}}  
\subsection{Photometric data\label{ss:photdata}}

%
% see the file :
% /disk-e/cretton/2ndvN2320/phot_observations/NGC2320.V.clean.imh 
%
NGC 2320 is a luminous elliptical galaxy with a heliocentric velocity of
5725 km/s, implying a distance of 76.3 $h_{75}$ Mpc. It has an apparent 
magnitude $m_V = 11.9$ (NED), which translates to an absolute magnitude of 
${\rm M}_V = -22.5$. 
%
% MGE model gives 12.2
%
A V-band image was obtained at the Steward Observatory $90''$
Telescope on March, 1996 with a 2k x 2k CCD and an exposure time of
600 seconds. The scale was 0.283 arcseconds per pixel. The image was
calibrated using published aperture photometry available via the
Hypercat catalogue (CRAL, Lyon), mostly consisting of data taken at
the OHP (Prugniel \& Heraudeau 1998). The effective radius $\Reff$ is
29.5$''$. Our photometric data is not good enough to detect a possible
central luminosity cusp (the seeing was about 1$''$) and no HST
archive image exists. In this paper, however, we cannot concentrate on
the core properties of NGC 2320, since we used a wide slit for
the spectroscopic observations.

\subsection{Stellar kinematic data\label{ss:kindata}}

\begin{figure*}[t!]
\epsfxsize=16.0truecm
 \epsfbox{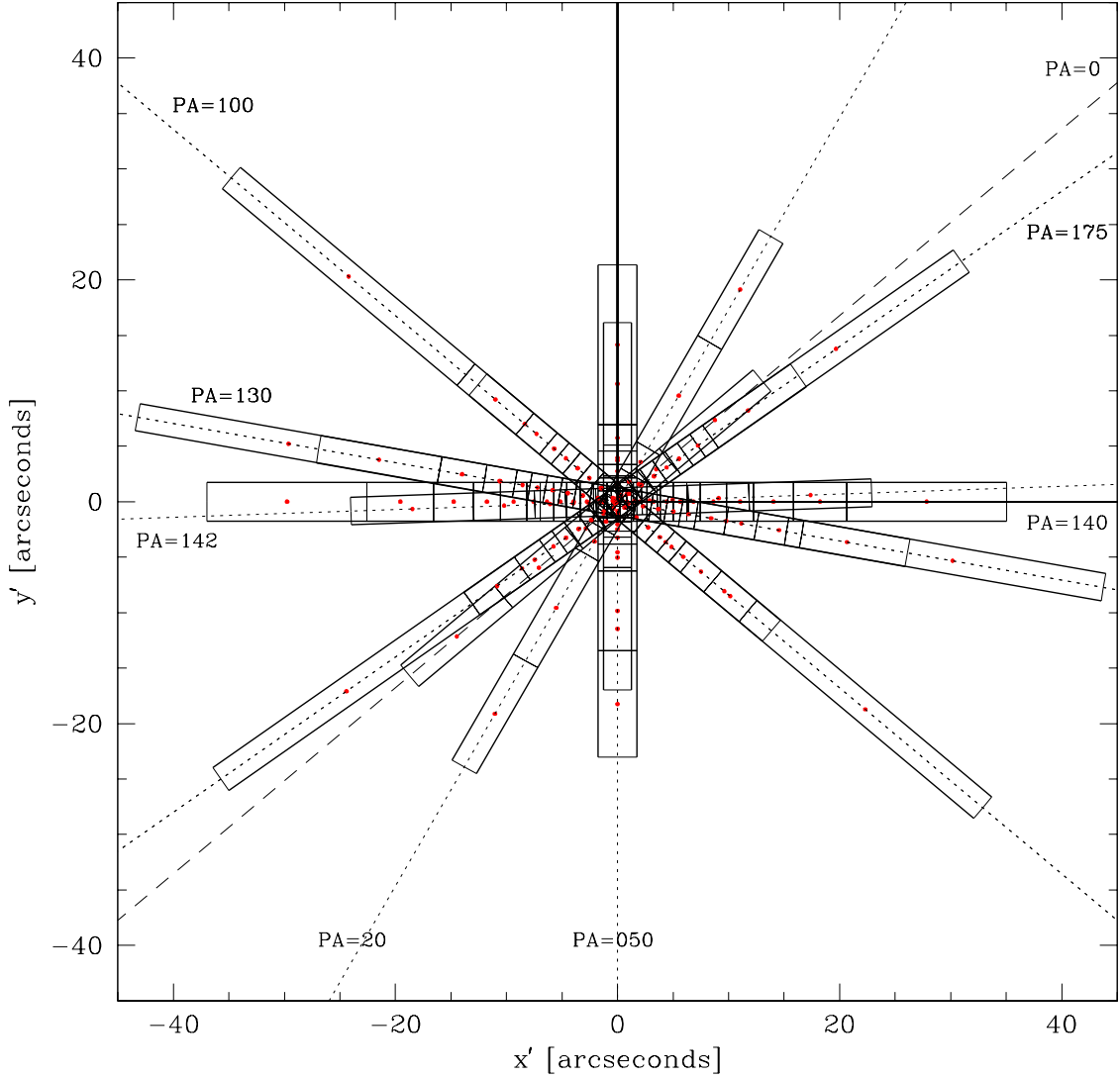}
 \caption[allslits]{Position of all the slits used to constrain 
 the dynamical models. The values of the various position angles are indicated
 (the major axis is PA=140 and the minor axis is PA=50). The effective 
 radius $\Reff$ is at $29.5''$. \label{fig:all_slits}}
\end{figure*}

NGC 2320 was observed during three different runs (February 96,
February 98 and March 1998) at the Multi Mirror Telescope (MMT) and
the 4 meter telescope at Kitt Peak.  Table~\ref{t:tbl2320-1} summarizes the
observations and Figure~\ref{fig:all_slits} illustrates all slit
positions on the plane of the sky.  All these various spectra are
statistically independent constraints even if they have the same PAs,
because they have a different radial extent (and binning) and, in some
cases, have been obtained at a different telescope. 

After standard reduction procedures with IRAF, the VPs were extracted
from the absorption lines and quantified using the Gauss-Hermite
decomposition (see e.g.  Rix \& White 1992, van der Marel \& Franx
1993). With the same notation as in Cretton \& van den Bosch (1999,
hereafter CvdB99), we parameterize these velocity profiles (VPs) using
Gauss-Hermite series with line strength $\gamma$, mean radial velocity
$V$, and velocity dispersion $\sigma$ as free parameters.  The
anisotropy is reflected in the shape of the VPs (and hence in the
values of the GH--moments $h_i$). Qualitatively, VPs more peaked than
a Gaussian tend to originate from radial anisotropy and translate into
greater $h_4$ values than their isotropic counterparts. Note that the
VP shapes not only depend on the dynamical structure, but also on the
gravitational potential, and can hence be used to break the
mass--anisotropy degeneracy (Dejonghe 1987, Merritt 1993, Gerhard
1993, see also Figure 2 in Gerhard \etal 1998, Carollo \etal 1995). In
total, we have 632 stellar kinematic constraints (158 data points
$\times$ 4 GH--moments $h_{1,\ldots,4}$, see Table~\ref{t:tbl2320-1}).

A careful look at the kinematic data
 (Figure~\ref{fig:fits_stellar_log}) reveals some {\it systematic}
 uncertainties, \eg the left/right assymetries. Also about half of the
 slits show a significantly higher ($\sim 40 \kms$) central dispersion
 value than for the remaining half. Given their formal errors ($\sim
 10 \kms$), such large differences amongst various PAs can not be
 explained through statistical fluctuations and are probably due to
 the strong gradients inside the central pixel, since we use a wide
 slit. Therefore we decide to replace the formal error bars of the
 central velocity dispersions by the variance amongst the (central)
 dispersion values from different slit positions.

\subsection{Kinematics of the ionized gas\label{ss:asymm}}

After subtraction of the broadened stellar template, emission was
found at 5007 [\AA] and 4965 [\AA], corresponding to the [OIII] lines
of ionized gas (see \eg Figure 2 of Rix \etal 1995).  However, only
the MMT data was of sufficient S/N to permit measurement of mean
velocity and velocity dispersion.

\begin{figure*}[t!]  
 \centerline{\epsfbox{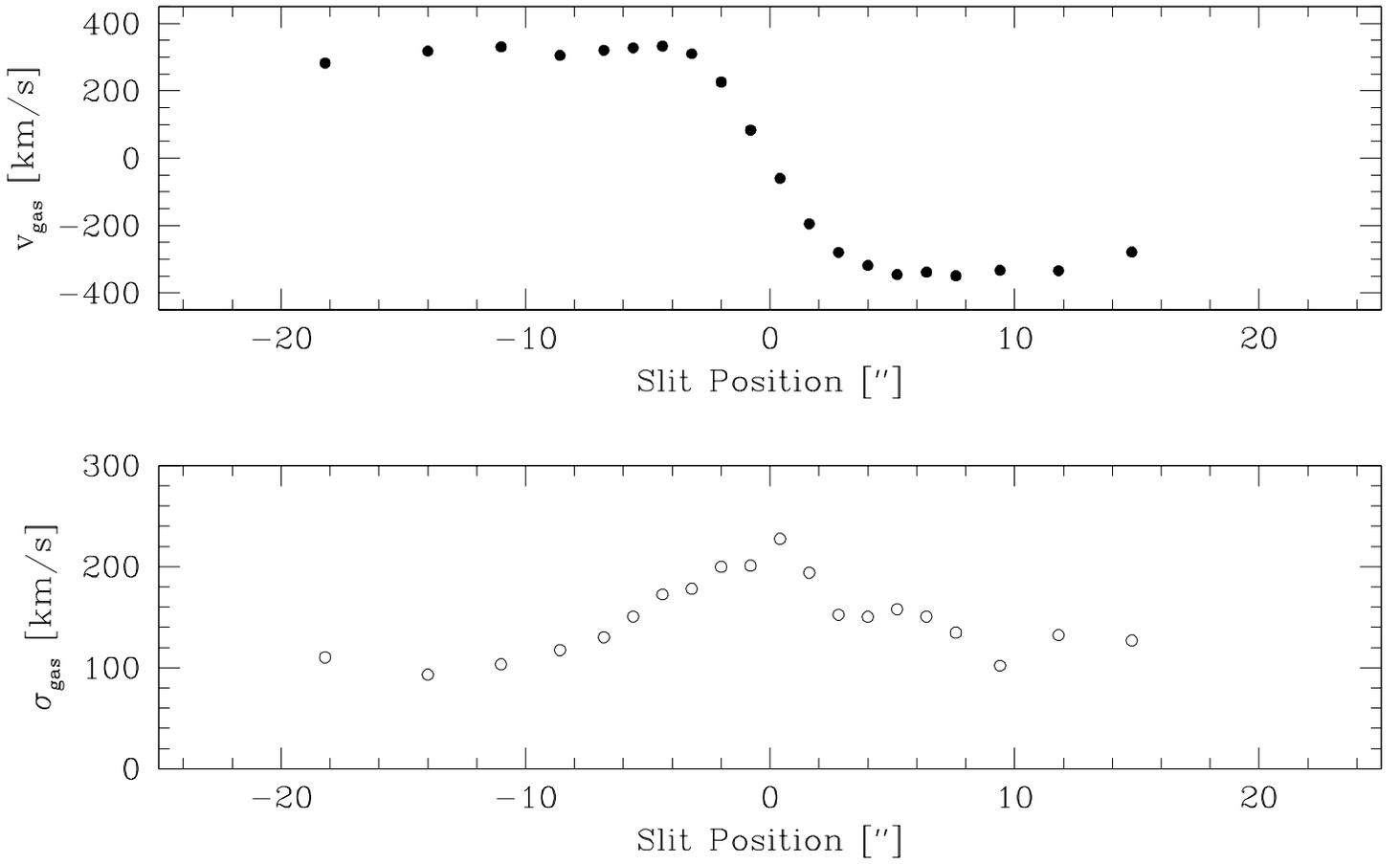}}
 \caption[v_and_sigma_gas]{Mean line--of--sight velocity and dispersion of 
 the gas along the major axis (MMT spectra). The central velocity 
gradient is affected by the wide slit ($3.5 ''$). 
\label{fig:v_and_sigma_gas}}
\end{figure*}

Usually, gas is assumed to rotate in the equatorial plane of a 
galaxy on nearly closed orbits. However, this is only true if the velocity
dispersion is negligible compared to the mean rotation velocity. In
Figure~\ref{fig:v_and_sigma_gas}, we show the mean line--of--sight
velocity and dispersion of the gas on the MMT major axis: The velocity reaches
325 km/s (inside $20''$) and the dispersion decreases roughly
exponentially from 220 km/s (center) to 100 km/s (at $15''$) and thus 
can not be neglected.

To derive the true circular velocity $v_{c}^{2}=R\;\partial\Phi_{\rm
 grav}/\partial R$ from the observed $v_\gas$ and $\sigma_\gas$, we
 proceed as follows (see \eg Neistein \etal 1999).  We first obtain
 the mean rotation velocity of the gas, $v_\phi(R)$, in the equatorial
 plane (\ie the plane of the disk) by de-projecting the observed
 $v_\gas$. Along the major axis, we have $v_\phi(R) = v_\gas/\sin \,i$,
 where $i$ is the inclination of the galaxy (away from face--on).
 Similarly, we have $\sigma_\phi(R) = \sigma_\gas/\sin \,i$. Note that
 in nearly edge--on disks, this is only approximately true, since the
 line--of--sight integration through the disk will reduce $v_\gas$
 relative to $v_\phi$. Neistein \etal (1999) estimated that this
 correction is less than 4\% for inclinations $i<70\deg$.

\begin{figure*}[t!]  
 \centerline{\epsfbox{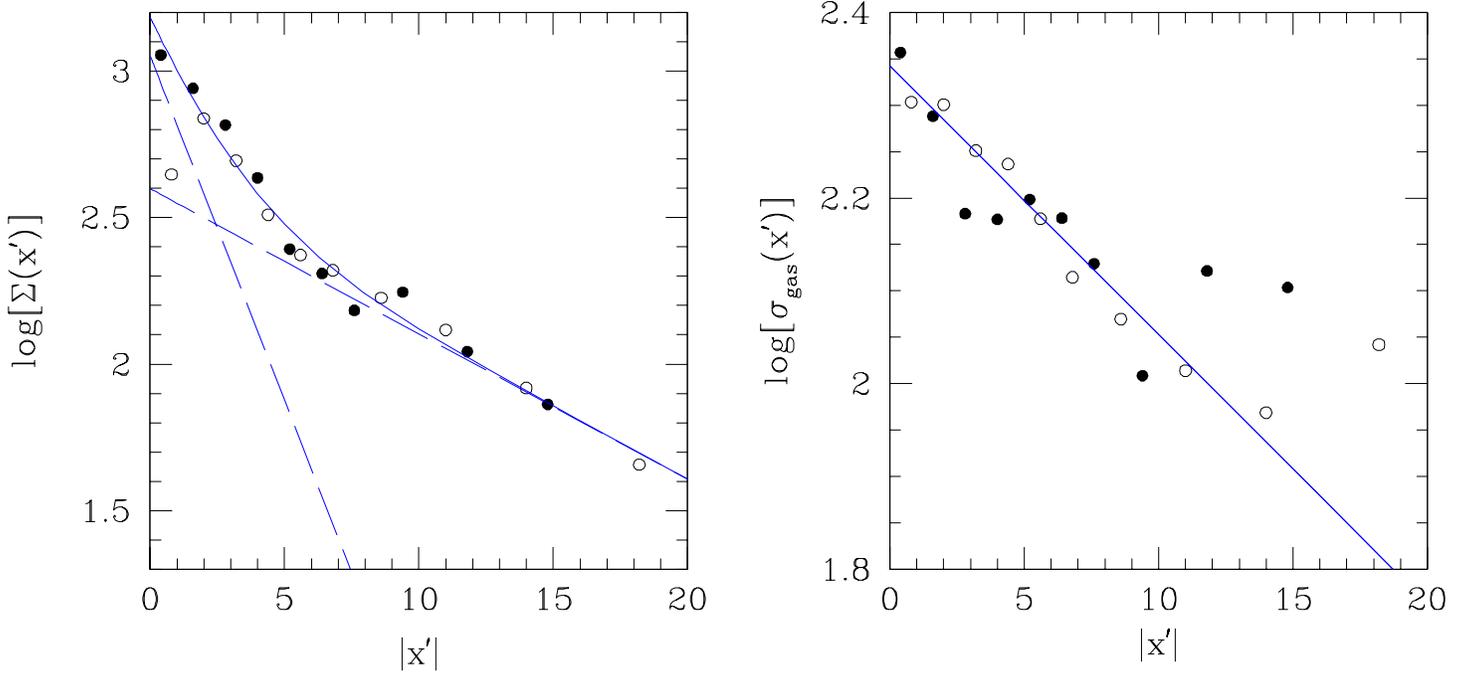}}
 \caption[gas_surf_and_disp]{Surface density profile of the ionized
 gas disk (left) and dispersion profile (right) along the major axis.
 The dots are the measured quantities, the full line is the fit. On
 the left panel, the dashed lines are the individual exponentials.
 Both sides of the major axis are shown: full symbols correspond to
 the positive side and open symbols to the negative side of the major
 axis.  
 \label{fig:gas_surf_and_disp}} 
\end{figure*}

\begin{figure*}[t!] 
 \centerline{\epsfbox{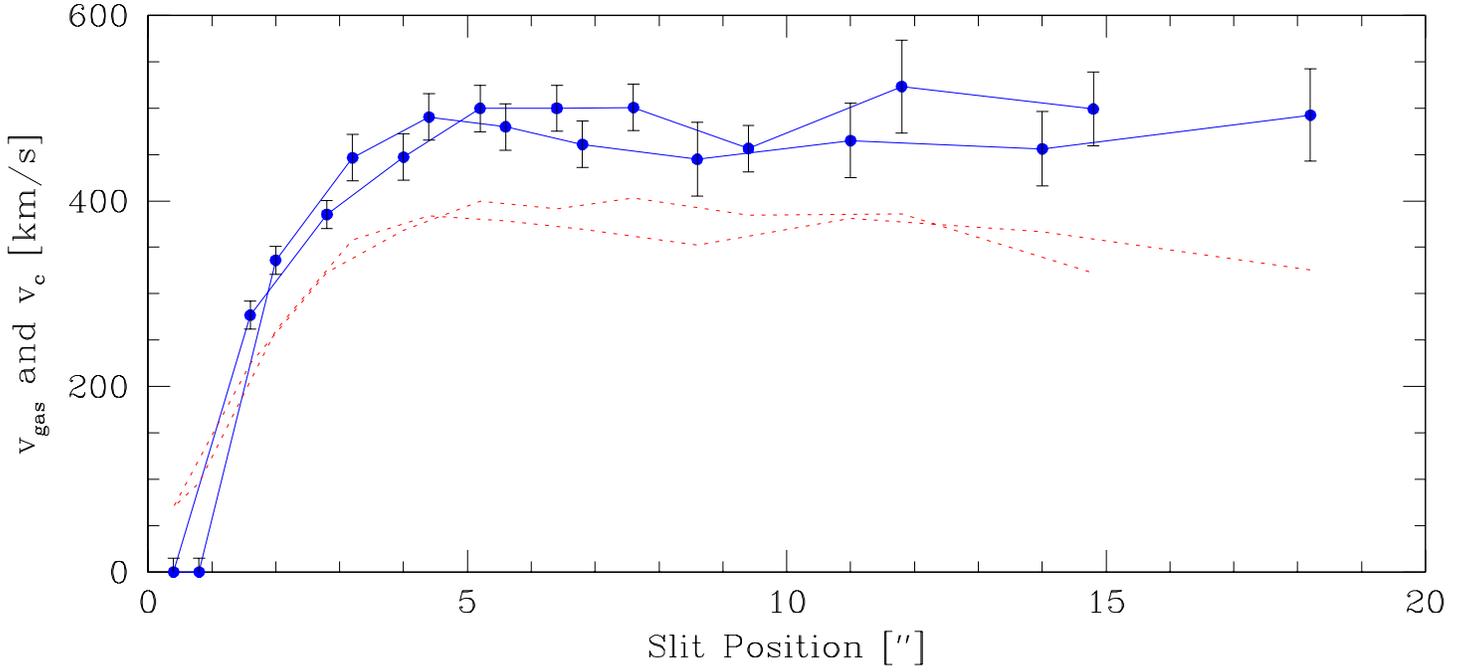}} \caption[vgas_and_vc]{Circular
 velocity as a function of radius (full line) when the asymmetric
 correction has been applied. For comparison, we plotted also the
 observed gas velocity (dotted line). Both sides of the major axis
 have been folded to the positive side. Only points at radii larger
 than $3''$ should be taken seriously, since during the correction we
 have assumed a flat rotation curve.  \label{fig:vgas_and_vc}}
\end{figure*}

%\pagebreak
We follow Binney \& Tremaine (1987, eq. 4--33) to obtain the circular 
velocity of the cold gas disk: 

\begin{equation}
\label{cold_disk}
v_{c}^{2} = \overline{v_{\phi}^{2}} + \sigma_{\phi}^{2} - \sigma_{R}^{2} -
{R \over \rho} {\partial(\rho\sigma_{R}^{2}) \over \partial R} - 
R {\partial(\overline{v_R v_z}) \over \partial z},
\end{equation}
where we assume that the final term can be neglected (\ie the gas
dynamics is close to isotropy). We fit the luminosity density of the
gas as $\rho(R) = \rho_1 \;\exp(-R/R_1) + \rho_2 \;\exp(-R/R_2)$ as
shown in the left panel of Figure~\ref{fig:gas_surf_and_disp}.  For a
flat rotation curve, the epicycle approximation gives
$\sigma_{R}^{2}(R) = 2 \sigma_{\phi}^{2}(R)$. Since the term
$\sigma_{R}$ enters a derivative in eq.~(\ref{cold_disk}) we find it
convenient to use an analytic expression instead of the (noisy) data
points, so we fit an exponential to $\sigma_\gas(R) = \sigma_0
\;\exp(-R/R_0) = \sigma_\phi(R)\;\sin \,i$ (right panel in
Figure~\ref{fig:gas_surf_and_disp}).  We can now evaluate the circular
velocity (Figure~\ref{fig:vgas_and_vc}). The error bars for $v_c(R)$
have been estimated from the left/right differences of the rotation
curve.

\section{The models\label{s:models}}
\subsection{The mass model\label{ss:mass_model_mge}}

As in CvdB99, we have used the Multi Gaussian
Expansion (MGE) of Emsellem \etal (1994) to construct a mass model for
NGC 2320. Briefly, in this formalism the surface brightness profile,
the mass density distribution and the PSF are all described as a sum
of Gaussian components. Free parameters include the center of each
Gaussian, its position angle, flattening, central intensity, and size
(\ie standard deviation) along the major axis.

We find that the density profile of NGC 2320 can be well fitted with 5
Gaussian components (see Table~\ref{t:tbl2320-2}), all with the same
position angle and center.  The mass density is expressed as
\begin{equation}
\label{MGE_density}
\rho(R,z) = \Upsilon\; \sum_{i=1}^{5} I_i \; \exp\Biggl[ -{1\over 2
\sigma_i^2}
\biggl( R^2 + {z^2 \over q_i^2} \biggr) \Biggr],
\end{equation}
where $\Upsilon$ is the mass--to--light ratio, $I_i$ is the central intensity
(in $\Lsun/{\rm arcsecond}^3$), $\sigma_i$ the standard deviation (in
arcseconds) and $q_i$ the flattening of each Gaussian. The
corresponding gravitational potential (and forces) are obtained by 
solving Poisson equation and can be written as 
one--dimensional quadratures (see CvdB99 for details). In this case, the 
mass distribution follows the light. 
 
\begin{figure*}[t!] 
 \epsfxsize=12.0truecm  
 \centerline{\epsfbox{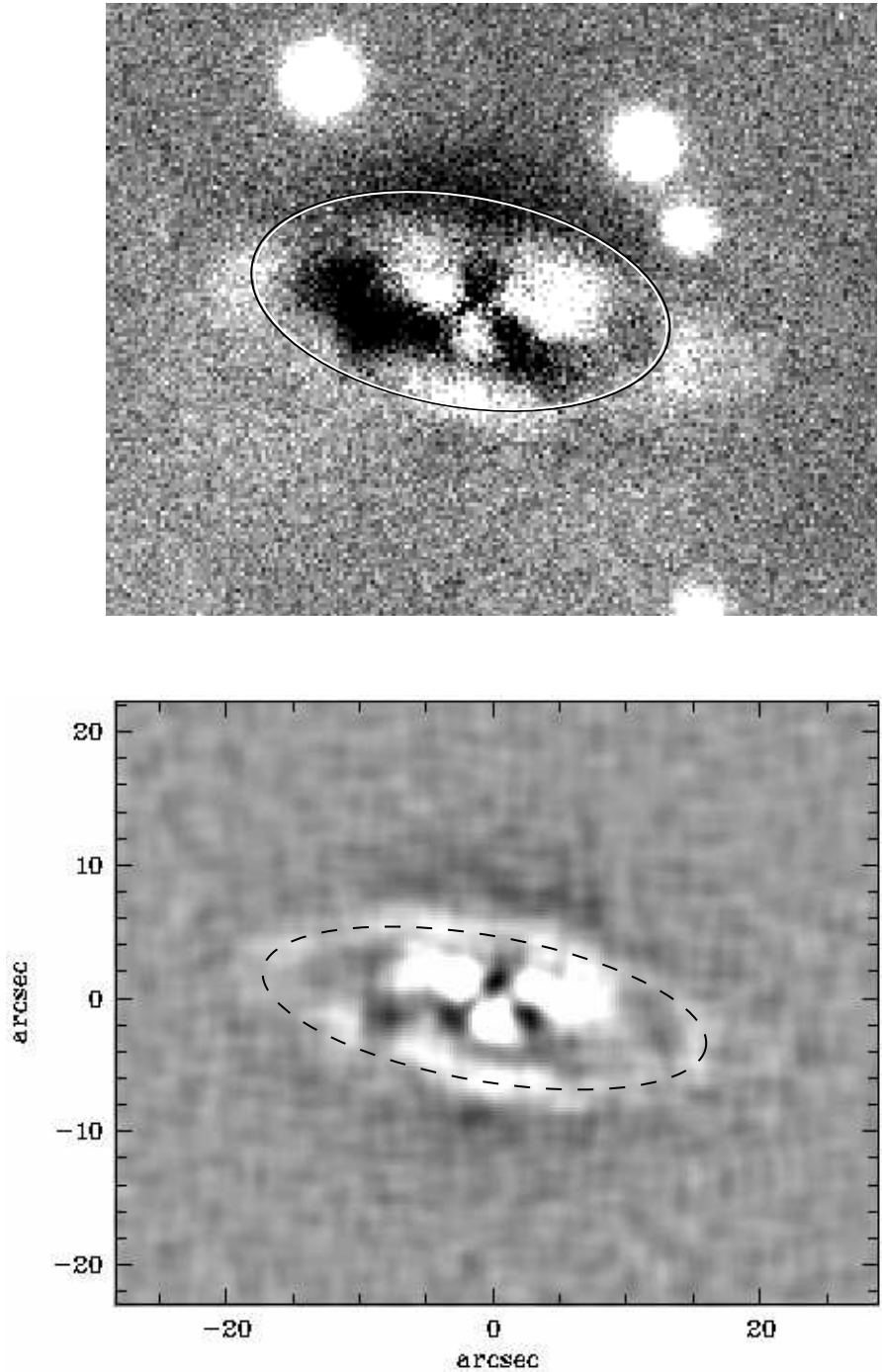}}
 \caption[Image_minus_model]{Ring--like structure
 once the MGE model has been subtracted from the image (top panel): an
 ellipse with an axis ratio of 0.5 is superposed (see text). The
 bottom panel shows an unsharp mask image with an ellipse of axis ratio
 of 0.32 (dashed line) that would translate into an inclination of
 $71\deg$. The big white dots outside the ring in the top panel are
 foreground stars.
 \label{fig:Image_minus_model}}
\end{figure*}

\begin{figure*}[t!]
 \centerline{\epsfbox{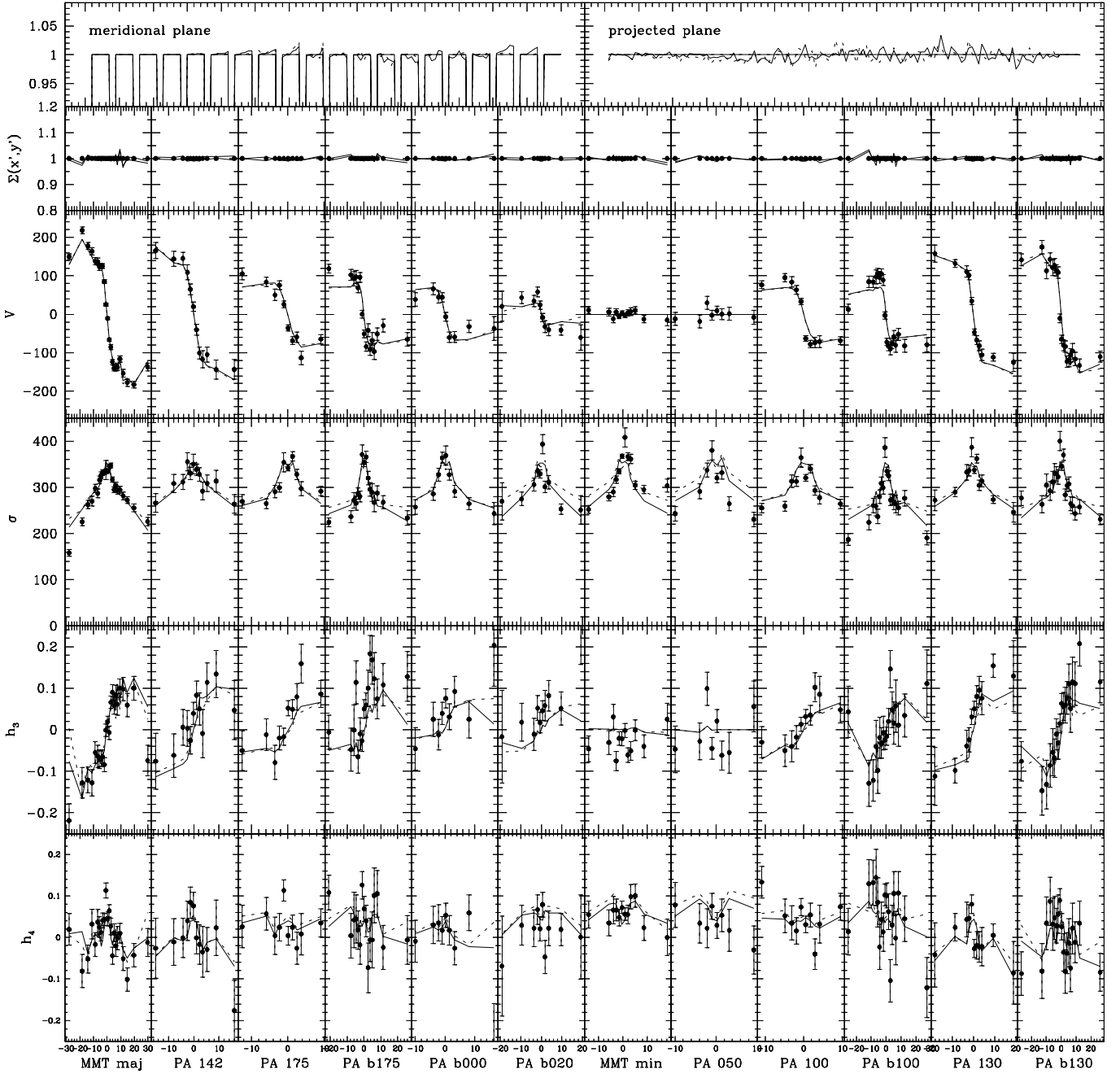}}
 \caption[fits_stellar_log]{Kinematic data for all the slits and fits
 of the two best fit models: the full line is the model where the mass
 follows the light and the dotted line is the model with the
 logarithmic potential. The columns have been organized in increasing
 Position Angle order, starting from the MMT major axis. The top line
 shows the fit of the (normalized) mass constraints in the meridional 
 plane and in the plane of the sky. In the fit of the meridional plane 
 constraints, we exclude the cells closest to the symmetry axis. This 
 explains the jagged appearance of the fit for these constraints.
 \label{fig:fits_stellar_log}} 
\end{figure*}

When we subtract the MGE model from the image, a disky structure
appears with a radial extent of $\sim 17''$. If we interpret this
feature as a ring of young stars and dust (hence darker on the near
side), we derive an axis ratio of 0.5 (top panel of
Figure~\ref{fig:Image_minus_model}). After additional filtering
(unsharp masking) a somewhat flatter ellipse of emission (with an axis
ratio of 0.32 corresponding to an inclination of $71\deg$)
appears (bottom panel of Figure~\ref{fig:Image_minus_model}). However,
our main results are not strongly dependent on the precise choice of
the inclination. For the remainder of this paper, we adopt an axis
ratio of 0.5, which translates into an inclination $i = 60\deg$.  The
apparent flattening of the galaxy (0.6--0.7) makes it unlikely to be
much more inclined, since galaxies intrinsically as flat as E6 are
very rare. Van den Bosch \& Emsellem (1998) found a similar stellar
ring in the nuclear region of NGC 4570.

We have explored another set of models in which the total gravitational 
potential and the luminous density are not related by Poisson 
equation: models with a logarithmic potential 
\begin{equation}
\label{Log_pot}
\Phi_{\rm log}(R,z) = {1 \over 2} \,v_{0}^{2} \;{\rm ln} 
\Biggl(R_{c}^{2} + R^2 + {z^2 \over q_{\Phi}^{2}} \Biggr).
\end{equation}

We still use the MGE density law (equation~\ref{MGE_density}) to
describe the luminosity profile of NGC 2320.  We have chosen the
simple logarithmic potential (with flat circular velocity at large
radii) to build a sequence of models with dark matter halos. We took
$q_{\Phi}=0.83$ such that the flattening of the corresponding mass
density profile is similar to that of the MGE models, $q_{\rho}\sim
0.5-0.6$. We adjust $R_{c}$ such as to match the inner parts of the circular
velocity curve and we scale the various models with $v_{0}$. The quantity 
$R_{c}$ is probably an upper limit of the true core radius, since we 
did not fit deconvolved data. Nevertheless it is not likely to have a 
large influence on the main results of this paper.

In Rix \etal (1997), we used dark halo profiles suggested by cosmological
N--body simulations (Navarro, Frenk, \& White, 1996) and adiabatically
contracted them to account for the luminous material. We found that
the best fit dark halo profile was nearly indistinguishable from a
simple logarithmic model over the radii probed by the kinematics of
integrated light.

\subsection{The dynamical models\label{ss:dynmodels}}

We construct dynamical models based on the orbit superposition
technique of Schwarzschild, described in detail in C99. Here, we limit
ourselves to listing the specific parameters we used for NGC 2320:
Starting with a "trial potential", $\Phi(R,z)$, we sample orbits on a
grid in integral space, energy $E$, vertical component of the angular
momentum $L_z$ and third integral $I_3$. We use 20 values of $E =
1/2\; R_c \; \partial \Phi/\partial R + \Phi(R_c,0)$, represented
through the radius of the circular orbit $R_c(E)$.  We space $R_c(E)$
logarithmically in $[0.1'',300.0'']$, which encloses more than 99 \%
of mass for our MGE model. For each $E$, we adopt 14 values of $L_z$
in $[-L_{z,{\rm max}},+L_{z,{\rm max}}]$ and 7 values of $I_3$ per
($E,L_z$) were adopted as in CvdB99.  We cannot assume that $I_3$ is
defined for every value of ($E,L_z$), \ie every orbit is
regular. Instead we sample starting points on the zero--velocity curve
(see C99); when the orbit is indeed regular, this starting point can
truly be interpreted as a third integral, $I_3$.

The orbit library is constructed by numerical integration of each
trajectory for a fixed amount of time, 200 periods of the circular
orbit at that E, using a Runge--Kutta scheme. During integration, we
store the fractional time spent by each orbit in a Cartesian
data--cube of observables $(\xp, \yp, \vlos)$, where $(\xp, \yp)$ are
the projected coordinates on the sky and $\vlos$ is the
line--of--sight velocity. This map is subsequently convolved with the
PSF and binned to match the various slit apertures. Furthermore, we
adopt logarithmic polar grids in the meridional plane $(R,z)$ and in
the $(\xp, \yp)$ plane with the same $R_c$--radial range and
sampling. Orbital occupation times are stored both on the intrinsic and
projected grids to make sure the final orbit model reproduces (to a
few percent accuracy) the MGE mass model. The lowest order velocity
moments of each orbit are also stored on such a grid to analyze the
dynamical structure of the resulting model (see C99 for details).

The mass on each orbit is determined from the non--negative
superposition of all orbits that best reproduces the kinematic data
within the errors and the projected and intrinsic MGE mass
profile. Using the NNLS algorithm (Lawson \& Hanson, 1974), smoothness
in integral space is enforced through a regularization technique. It
allows the derivation of smooth anisotropy profiles of a few selected
models.

\section{Results\label{s:resultstwentytwo}}
\subsection{Fits to the data\label{ss:fits_data}}

In Figure~\ref{fig:fits_stellar_log}, we show the two best fit models
to all the data, in the case where mass follows light (full line, see
Section~\ref{ss:mass_model_mge}) and in the logarithmic potential case
(dotted line). The first row displays the fit to the mass constraints
(intrinsic and projected), normalized to unity. The second row shows
the fit to the integrated surface density in the slit bins (also
normalized to unity).  The subsequent rows show the kinematic data and
the fits of both models.  Each column corresponds to a different
position angle on the sky (see Table~\ref{t:tbl2320-1}).
  
Both the MGE and the isothermal model fit the data comparatively well.
The differences appear at large radii (\eg in the velocity
dispersions), where the circular velocities of the two models (\ie the
enclosed masses) start to diverge. Both models have problems
fitting about half of the central dispersion points. This may be due
to the absence of a density cusp (and/or central black hole) in our
MGE model, but as mentioned earlier (see Section~\ref{ss:kindata}),
there are systematic problems with the data in the center. Moreover,
for some position angles, the observed dispersion seems to drop too
fast compared to the models, whereas for other position angles,
 it is well fitted. 

\subsection{Goodness of fit\label{ss:stats}}

We wish to assess statistically which range of $\Upsilon_V$ provides an 
acceptable fit to the data. We can construct model sequences with
different $\Upsilon_V$ by simply rescaling one orbit library. We then
perform a NNLS fit for each new model and compute the $\chi^2$
($\Upsilon_V$). In this way, we can study the stellar $\chi^2$
distribution, $\chi^2_{\rm stars}$, as a function of $\Upsilon_V$. A
similar $\chi^2$ comparison is done with the gas data: After rescaling
a model by $\Upsilon_V$, we ask by how much the new circular velocity
(rescaled by $\sqrt{\Upsilon_V}$) differs from the one derived from the
gas measurements (see Section~\ref{ss:asymm}), yielding a distribution
$\chi^2_{\rm gas}$.  These two distributions can be combined to get
better constraints on $\Upsilon_V$. The combined probability
distribution is ${\cal P}(\Upsilon_V) \sim {\rm exp}[-\chi^2_{\rm
comb}] = {\rm exp}[-(\chi^2_{\rm stars} + \chi^2_{\rm gas})]$
(Press \etal 1992).

\begin{figure*}[t!]
 \centerline{\epsfbox{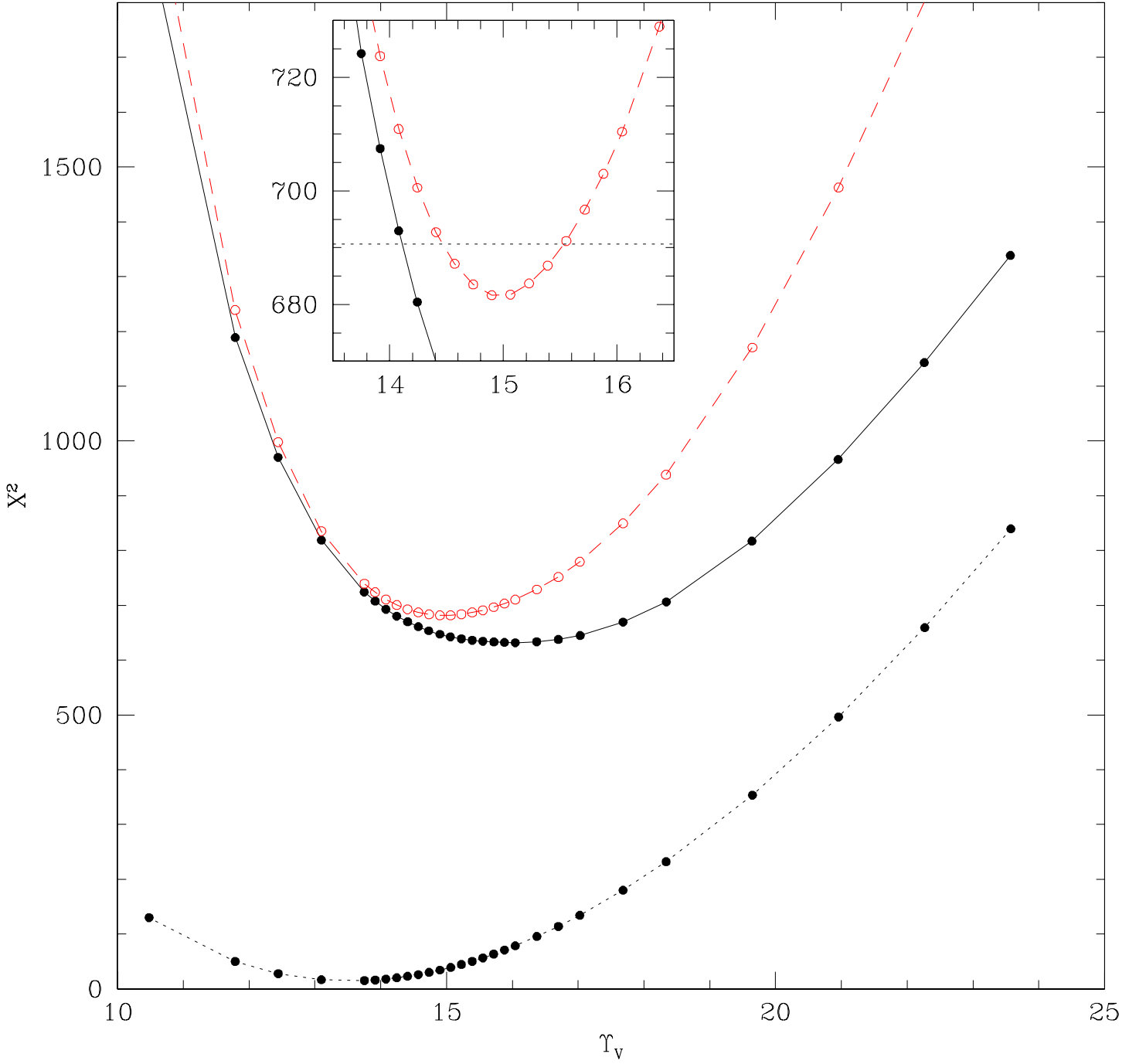}}
 \caption[chi2_comb]{$\chi^2$ distributions
 for the models with spatially constant $\Upsilon_V$.  The dotted line is
 $\chi^2_{\rm gas}$, the full line is $\chi^2_{\rm stars}$ and the
 dashed line (with open dots) corresponds to the combination of both
 type of constraints. The dots represent the actual model
 calculations. The inset shows in more details the region near the
 (combined) $\chi^2$ minimum. The horizontal dotted line has $\chi^2 =
 \chi^2_{\rm min}+9$.
 \label{fig:chi2_comb}}
\end{figure*}

\begin{figure*}[t!]
 \centerline{\epsfbox{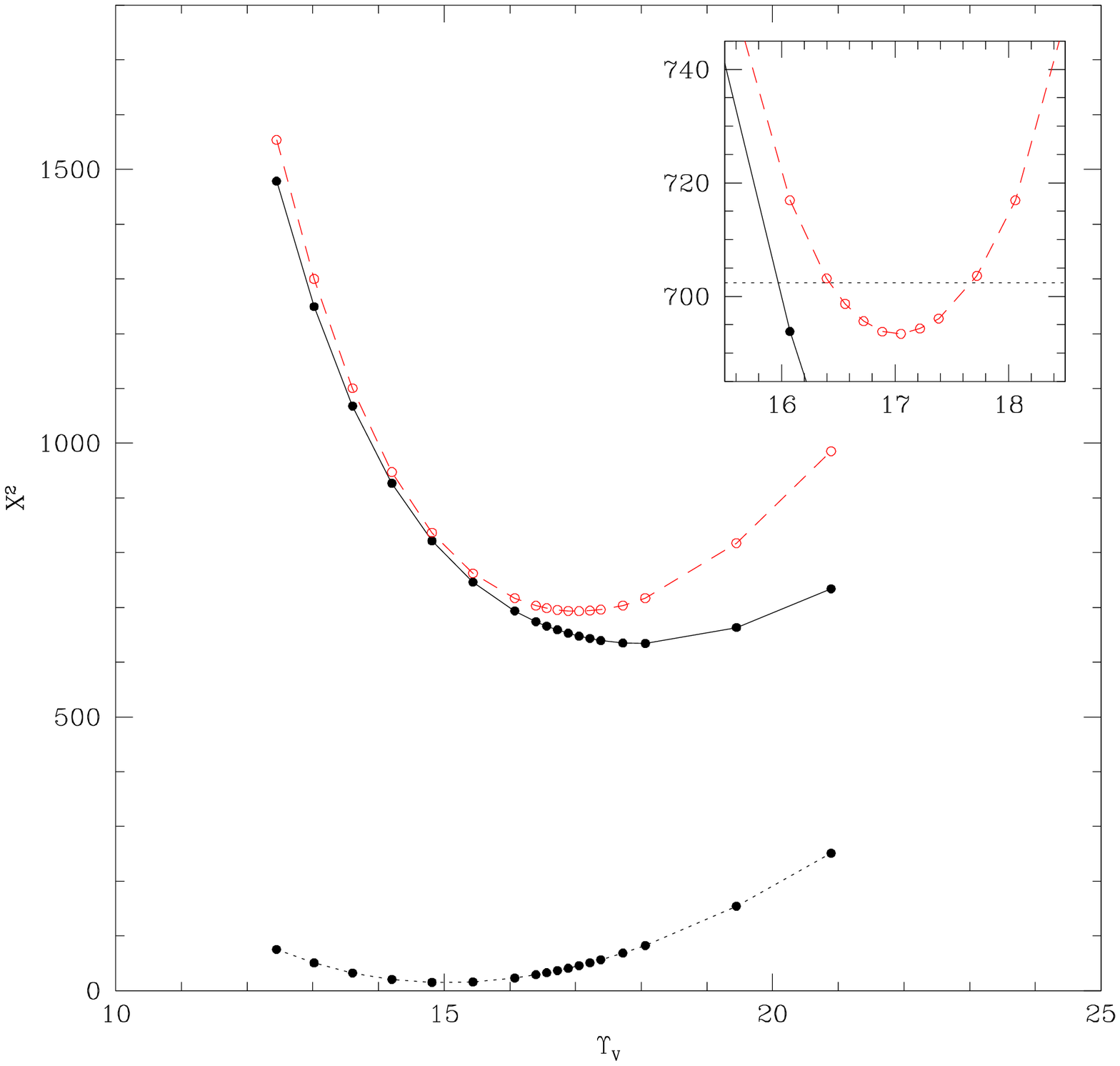}}
 \caption[chi2_comb_LOG]{Same as Figure~\ref{fig:chi2_comb}, but for 
 the model with the logarithmic potential.
 \label{fig:chi2_comb_LOG}}
\end{figure*}

For an acceptable fit, one expects $\chi^2$ to be roughly equal to the
number of data points ${\rm N_{data}}$ minus the number of degrees of
freedom ${\rm N_{DOF}}$, which is not the case in the stellar fits
because the statistical error bars do not reflect all uncertainties
(see Section~\ref{ss:kindata} and Figure~\ref{fig:fits_stellar_log}).
We show in Appendix A that the effective ${\rm N_{DOF}}$ is much
smaller that ${\rm N_{data}}$.  Following Kochanek (1994), we rescale
the $\chi^2_{\rm stars}$ distribution such that $\chi^2_{\rm stars,\,
min} = 632$.  Similarly, we consider the 15 gas data points with
$|\xp|>3''$ for which the gas rotation curve is roughly flat and
rescale the $\chi^2_{\rm gas}$ distribution, so that $\chi^2_{\rm
gas,\, min} = 15$.  In Figure~\ref{fig:chi2_comb}, we plot
$\chi^2_{\rm gas}$, $\chi^2_{\rm stars}$, and the combined
$\chi^2_{\rm comb}$ distribution as function of $\Upsilon_V$ for the
models in which the mass follows the light. The minimum $\chi^2_{\rm
min,\,comb}$ is attained for $\Upsilon_V = 15.0\pm 0.6$, where the
error bar corresponds to the formal 99.73 \% confidence level or
$3\,\sigma$ interval (\ie the range of $\Upsilon_V$ for which
$\chi^2_{\rm comb} = \chi^2_{\rm min} + 9$).  We perform the same
exercise for the model with the logarithmic potential.  In that case,
$\Upsilon_V$ is defined as the mass enclosed within $15''$ divided by
the total amount of light in the same volume, and we find $\Upsilon_V
= 17.0\pm 0.7$ at the same confidence level
(Figure~\ref{fig:chi2_comb_LOG}).

In order to compare with the sample of 37 bright ellipticals studied by
van der Marel (1991), we first calculate the total absolute magnitude
of NGC 2320 in B using his choice for $H_0$, and find ${\rm M_B} = -23
\;h_{50}$. We then translate our value of $\Upsilon_V = 15$ in the V--band
into the R--band to place NGC 2320 on van der Marel's Figure 6 (upper
left panel). Using V--R of 0.77 (typical for K0 giants) we find
$\Upsilon_R = 8.0$. This value is 1.8 times higher than the prediction
of his least--square fit relation at that absolute magnitude, making
NGC 2320 a slight outlier. We have used data from larger radii than in 
van der Marel's analysis and this could account for some of the
discrepancy. Furthermore radially anisotropic models like ours tend to
yield higher mass--to--light ratio than isotropic models
(see Figure 6 of van der Marel \& Franx, 1993); however, this would decrease
$\Upsilon_R$ by only $\sim 20 \%$ for NGC 2320.

\subsection{Intrinsic velocity dispersions\label{ss:dispersions}}

\begin{figure*}[t!]
 \centerline{\epsfbox{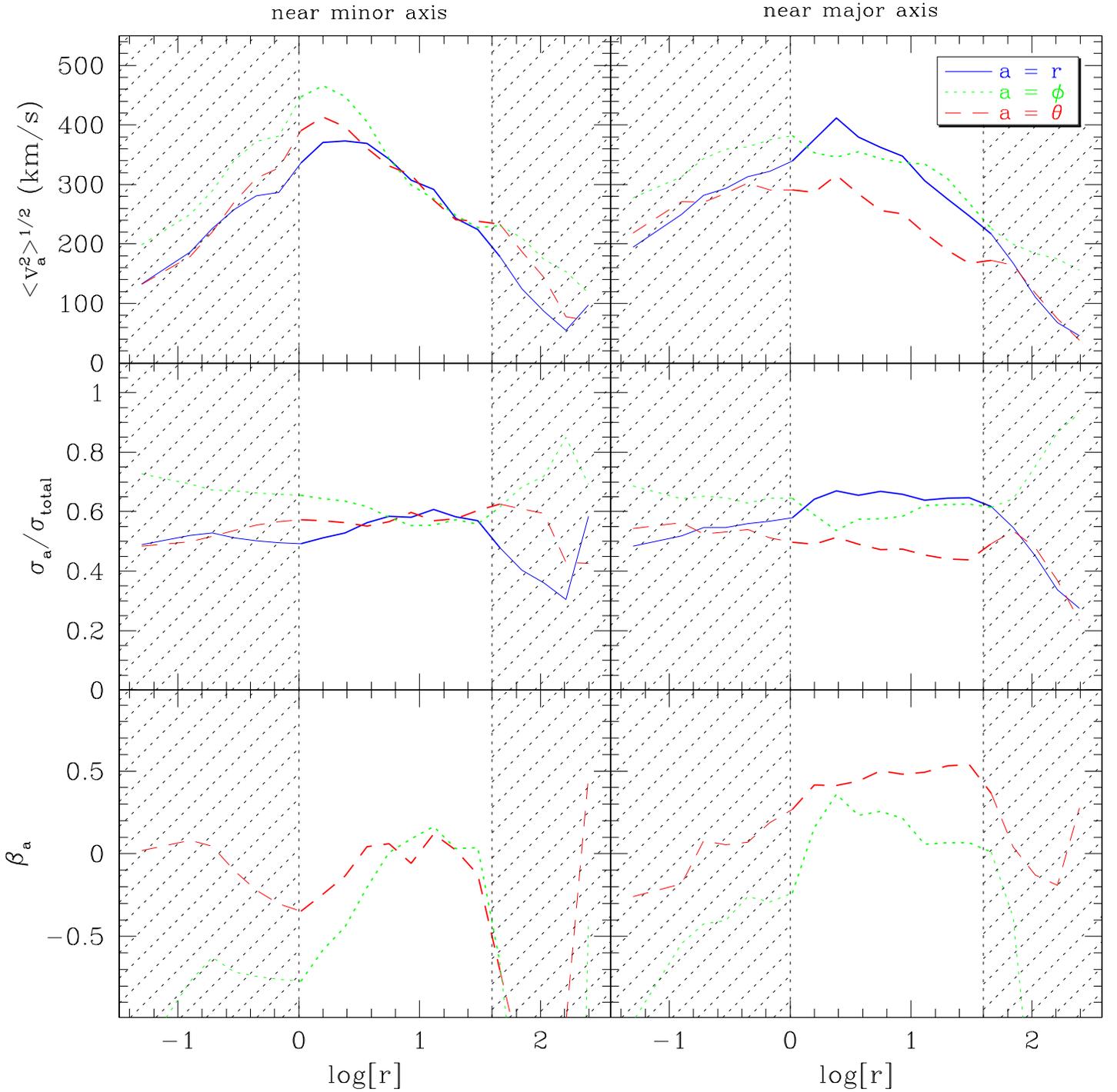}}
 \caption[disp_ml_15]{Anisotropy profiles of
 the model with $\Upsilon_V$ = 15. The first column is an average around the
 symmetry axis, and the second one around the equatorial plane. The
 first row shows the second moments in km/s, the second one the ratio
 of the dispersions to the total dispersion, and the last one shows the
 anisotropy parameter defined as $\beta_a =
 1-\sigma_{a}^{2}/\sigma_{r}^{2}$. The inset in the upper right panel
 explains the line convention.
 \label{fig:disp_ml_15}}
\end{figure*}

\begin{figure*}[t!]
 \centerline{\epsfbox{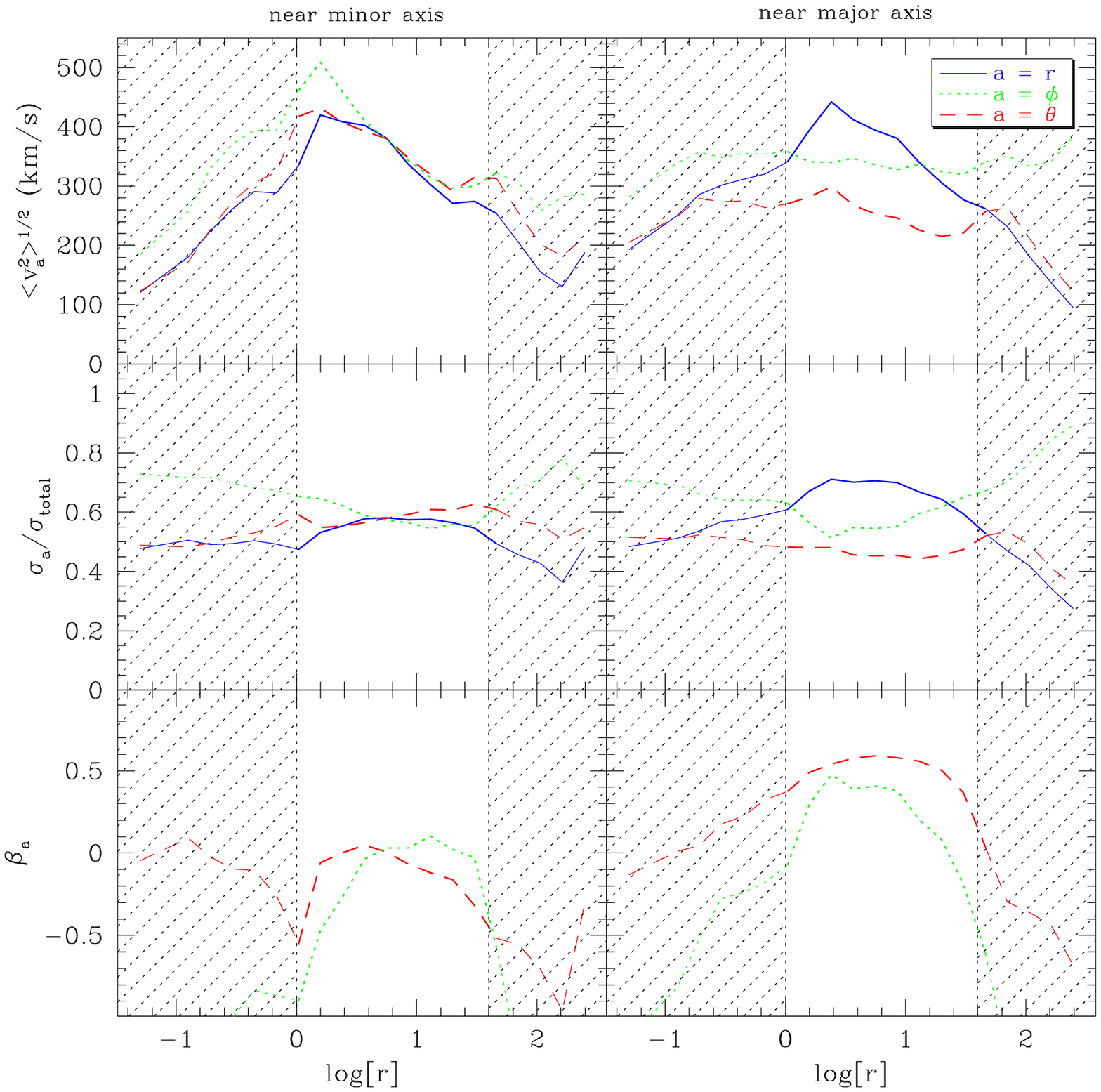}}
 \caption[disp_17_LOG]{Same as Figure~\ref{fig:disp_ml_15},
 but for the model with the logarithmic potential. \label{fig:disp_17_LOG}}
\end{figure*}

Using solutions with a statistically indistinguishable regularization
(as defined in Section 5.5 of CvdB99), we compute second moments
$\langle v_{R}^{2} \rangle^{1/2}, \langle v_{\phi}^{2}\rangle^{1/2},
\langle v_{\theta}^{2} \rangle^{1/2}$, ratio of dispersion profiles 
$\sigma_R, \sigma_{\phi}, \sigma_{\theta}$, and the anisotropy
parameter~$\beta$ (see \eg Binney \& Tremaine, 1987) along the major
and minor axis for the best fit models.  In
Figure~\ref{fig:disp_ml_15} we show these quantities for the best fit
MGE model with $\Upsilon_V$ of 15. At a given radius the results are
averaged over cells with different polar angles $\theta$: in the
meridional plane, the first three angular sectors closest to the minor
axis are averaged together (left column of
Figure~\ref{fig:disp_ml_15}), as well as the remaining four closest to
the major axis (right column).  The first row displays the second
moments as a function of radius. These moments are normalized by the
total dispersion in the second row. In the last row, we plot the
anisotropy parameter $\beta_a = 1-\sigma_{a}^{2}/\sigma_{r}^{2}$ for
$a=\phi$ (dotted line) or $a=\theta$ (dashed line). Positive $\beta$
corresponds to radial anisotropy.  Figure~\ref{fig:disp_17_LOG} shows
the same, but for the model with the logarithmic potential.  For NGC
2320 the anisotropy exhibits similar general behavior for both types
of models: they are dominated by radial anisotropy for most radii
constrained by the data (between the shaded areas in
Figures~\ref{fig:disp_ml_15} and \ref{fig:disp_17_LOG}). On the major
axis, $\beta_\theta \simeq 0.4 - 0.5$ and $\beta_\phi
\simeq 0.0 - 0.3$, whereas $\beta_\theta \sim \beta_\phi \sim 0.0$ on
the polar axis (except in the central $3''$). 

\begin{figure*}[t!]
 \centerline{\epsfbox{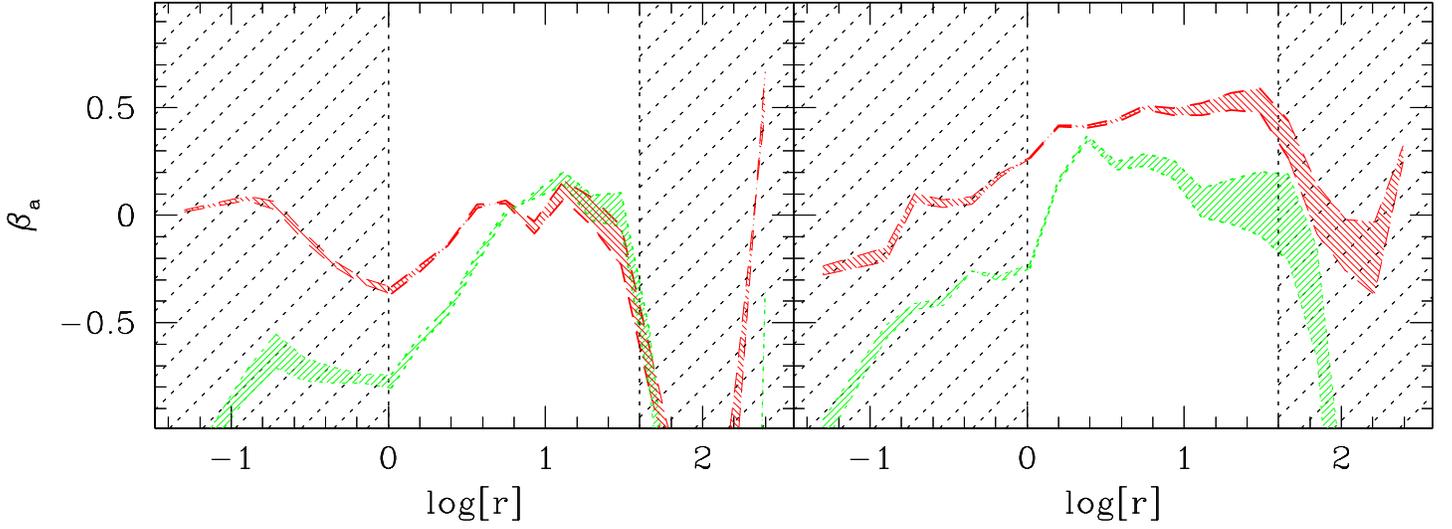}}
 \caption[betas_with_boundary] {The
 anisotropy parameter as in Figure~\ref{fig:disp_ml_15}: the densely shaded
 region indicates the values of $\beta$ for models which lie in the
 formal 99.73 \% confidence level region around the best fitting model
 (see Figure~\ref{fig:chi2_comb}). \label{fig:betas_with_boundary}}
\end{figure*}

\begin{figure*}[t!]
 \centerline{\epsfbox{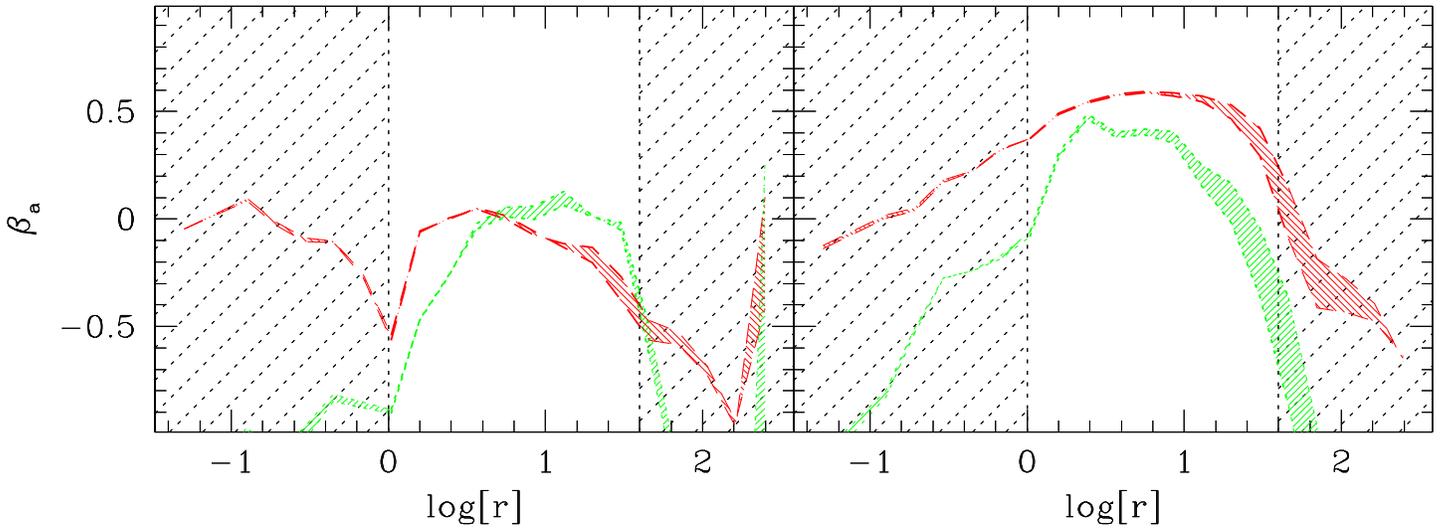}}
 \caption[betas_LOG_with_boundary] {Same as Figure~\ref{fig:betas_with_boundary},
  but for the logarithmic potential mass model.
 \label{fig:betas_LOG_with_boundary}}
\end{figure*}

In Figure~\ref{fig:betas_with_boundary} and
\ref{fig:betas_LOG_with_boundary}, we concentrate on the anisotropy 
parameter, $\beta$, to explore how much it varies among different
statistically acceptable models.  We compute the $\beta$--profiles of
the two models that correspond to the 99.73 \% confidence limit and
shade the region in between them (Figure~\ref{fig:betas_with_boundary}
and Figure~\ref{fig:betas_LOG_with_boundary}), effectively estimating
an error bar for $\beta$.  All models in the 99.73 \% confidence
interval follow the same trend: they are radially anisotropic on the
major axis and more nearly isotropic on the minor axis.

\section{Conclusions and Discussion\label{s:conclusionstwentytwo}}

We have constrained the orbital anisotropy of the stars in NGC 2320 by
combining stellar VPs along multiple position angles with the
kinematics of the ionized emission line gas. The nearly
two--dimensional coverage of the galaxy by many long slit spectra
tightly constrains the stellar dynamical model, while the gas
kinematics provides constraint on the normalization of the
gravitational potential, independent of stellar anisotropy issues.

The dynamical models are constructed following our extension of
Schwarzschild's orbit superposition technique described in C99. The
best fitting model is dominated by radial anisotropy near the
equatorial plane and is more isotropic on the polar axis. This
conclusion is valid both for models in which the mass is proportional
to the light and for models with a flat rotation curve at large radii
(logarithmic potential).  We find a best fit $\Upsilon_V$ of $15.0
\pm 0.6$ for the radially constant $\Upsilon_V$ model and
$\Upsilon_V(<15'')=17.0\pm 0.7$ for the logarithmic one.  We study the
dispersion profiles of models for these intervals of $\Upsilon_V$ in
order to determine the range of anisotropy among models that all fit
the data. This range is small in general and all models in the
$\Upsilon_V$ interval follow the same trend (\ie radial anisotropy in
the equatorial plane, isotropy near the symmetry axis).  Radial
anisotropy of the stellar objects has also been inferred from the
modelling of some other objects (NGC 2434, NGC 1600, NGC 6703, NGC
1399, NGC 3377, NGC 3379, NGC 4473, NGC 5845 and M87).

It seems worth comparing to the dynamical structure of merger remnants
produced by N--body simulations.  Many merger simulations leading to
the formation of an elliptical galaxy have concentrated on mergers of
a pair of disk galaxies (Barnes 1992; Barnes \& Hernquist 1996), of
several disk galaxies (Barnes 1989; Weil \& Hernquist 1996) or of
small virialized clusters of spherical galaxies (Funato, Makino \&
Ebisuzaki 1993; Garijo, Athanassoula \& Garcia--Gomez 1997). Dubinski
(1998) carried out simulations embedded in a cosmological context and found
that the anisotropy of the brightest cluster galaxy grows gently from
$\beta=0.0$ near the center to $\beta=0.5$ at 3 $\Reff$. Inside one
$\Reff$, his N--body model is mildly radially anisotropic ($\beta <
0.3$). This is qualitatively consistent with our results for NGC 2320.

A comparison of NGC 2320 with Barnes' results is more complicated: In
his collisionless simulations, Barnes (1992) typically finds {\it
triaxial} merger remnants with a large fraction of box orbits (see \eg
his figure 20). These box orbits do not exist in the axisymmetric case
where only tube orbits are present. Therefore our axisymmetric models
have no other choice than to put a lot of weight on low $L_z$ tubes to
produce radial anisotropy. This is not the case for Barnes' models:
the box orbits (with $L_z=0$) are responsible for the radial
anisotropy whereas his population of tube orbits seems fairly uniform
in $L_z$ (see his Figures 20b, 21b and 22b).

Black holes, steep central density cusps and central gas
concentrations can efficiently scatter stars on box orbits and destroy
the triaxiality (see \eg Gerhard \& Binney 1985; Barnes \& Hernquist
1996; Merritt \& Quinlan 1998). In this scenario, the galaxy shape
evolves towards axisymmetry, and its DF becomes closer to a
$f(E,L_z)$--form (Merritt 1999). Therefore, our {\it anisotropic}
results for NGC 2320 argue against a scenario in which large amounts
of gas flowed to the center during its formation. It would be
interesting to investigate the existence of a central black hole and
of a steep stellar cusp in NGC 2320, using higher resolution
observations.

%------------------------------------------------------------------------
% Acknowledgments

\section*{Acknowledgments}
  
We thank Eric Emsellem for discussions and help with the MGE software
and Roeland van der Marel for a careful reading of the manuscript.
N.C. thanks the Max--Planck Institut f\"ur Astronomie for hospitality
during the spring of 1999, when the bulk of this work was done.

%------------------------------------------------------------------------
\clearpage

\appendix

\section{Estimation of the number of degrees of freedom}
\label{s:App1}

For a good $\chi^2$ fit, one expects on average $\chi^2$ to be equal
to $\chi^2_{\rm expected} = {\rm N_{data} - N_{DOF}}$, where ${\rm
N_{data}}$ is the number of data points and ${\rm N_{DOF}}$ is the
number of degrees of freedom, if the errors are properly estimated and
normally distributed. In our Schwarzschild models, we have many more
weights that can be adjusted than we have data constraints. Yet, the
projected properties of our orbits are not linearly independent and we
have additional non--negativity, self--consistency and symmetry
constraints; hence ${\rm N_{DOF}}$ is considerably smaller than the
number of orbits. The key question is therefore how to determine the
effective $\rm N_{DOF}$ for our orbit models ?

\begin{figure*}[t!]
 \centerline{\epsfbox{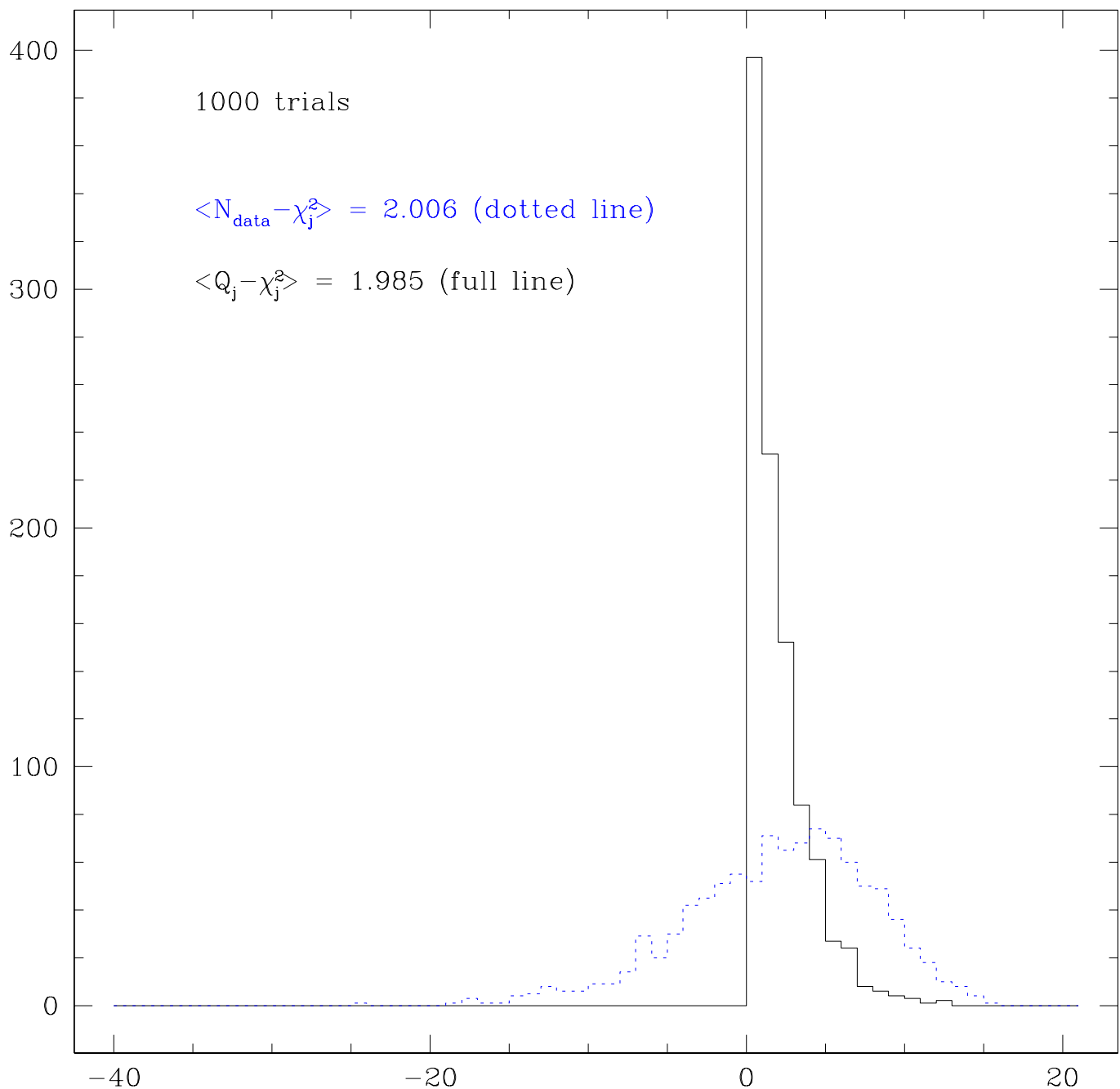}} \caption[straight_line]
 {Histograms showing the broad distribution ${\rm N_{\rm data} -
 \chi^2_j}$ (dotted line) and the narrower distribution $Q_j -
 \chi^2_j$ (full line) for 1000 Monte--Carlo realizations. The
 effective $\rm N_{DOF}$ is the mean of these distributions. Here
 we show the case of a straight line, \ie $\rm N_{DOF}=2$.
 \label{fig:straight_line}}
\end{figure*}

 One can estimate $\rm N_{DOF}$ as $\langle {\rm N_{\rm data}} -
 \chi^2_j \rangle$, where this average is done over many Monte--Carlo
 realizations of the data and where $\chi^2_j$ of the $j$-realization
 is defined as $\sum_i^{\rm N_{data}} (y_i - y_{i, {\rm best-fit}})^2
 / \sigma_i^2$. Each of these realizations is generated according to a
 Gaussian distribution around each data point, with a variance equal
 to the corresponding error bar. The distribution ${\rm N_{data}} -
 \chi^2_j$ is broad, so its mean is not accurately defined for a
 modest number of realizations. Therefore we prefer to estimate $\rm
 N_{DOF}$ as the mean $\langle Q_j - \chi^2_j \rangle$, where $Q_j =
 \sum_i^{\rm N_{data}} (y_i - y_{i, {\rm true}})^2 / \sigma_i^2$ for
 the $j^{th}$-realization. The $y_{i, {\rm true}}$ are the true
 underlying data points.  In Figure~\ref{fig:straight_line}, we
 illustrate the estimation of $\rm N_{DOF}$ with 20 fake data points
 drawn from and then fitted to a straight line function, i.e. $\rm
 N_{DOF}$=2. We generate 1000 data sets and show that the ${\rm N_{\rm
 data}} - \chi^2_j$ distribution is broader than $Q_j - \chi^2_j$.

We can apply the same procedure to the more complex case of our orbit
 model for NGC 2320.  In the presence of the systematic deviations in
 the NGC 2320 data, we took the best--fitting model to the
 observational data as the true underlying distibution and generated
 10 Monte--Carlo data sets normally distributed around it (according
 to the observed error bars). For each of these models, we performed a
 NNLS fit and computed $\langle Q_j - \chi^2_j \rangle$. We found $\rm
 N_{DOF} \simeq 25$, much smaller than ${\rm N_{\rm data}}$, which is
 632 in this case. Therefore we have about 25 times more data
 constraints than effective degrees of freedom, implying a well--posed
 fitting problem.

% --- References

\section*{References}
\begin{description}
\setlength{\itemsep}{0pt}
\setlength{\labelsep}{0pt}
\setlength{\parsep}{0pt}
\parskip=0pt

\item Barnes, J. E. 1989, Nature, 338, 123

\item Barnes, J. E. 1992, ApJ, 393, 484

\item Barnes, J. E., \& Hernquist, L. 1996, ApJ, 471, 115
   
\item Binney, J.~J., \& Tremaine S. D. 1987, Galactic Dynamics
                  (Princeton: Princeton University Press)       

\item Carollo, C. M., de Zeeuw, P. T., van der Marel, R. P., 
Danziger, I. J., \& Qian, E. E. 1995, ApJ, 441, L25

\item Cinzano, P., van der Marel, R. P. 1994, MNRAS, 270, 325

\item Cretton, N., de Zeeuw, P. T., van der Marel, R. P., \& Rix,
	H.-W. 1999, ApJS, 124, 383 (C99)
 
\item Cretton, N. \& van den Bosch, F. C. 1999, ApJ, 514, 704 (CvdB99)

\item Dejonghe, H., de Bruyne, V., Vauterin, P., Zeilinger, W. W. 1996, 
A\&A, 306, 363

\item Dejonghe, H. 1987, MNRAS, 224, 13

\item Dubinski, J. 1998, ApJ, 502, 141

\item Emsellem, E., Monnet, G., \& Bacon, R. 1994, A\&A, 285, 723
 
\item Funato, Y., Makino, J., \& Ebisuzaki, T. 1993, PASJ, 45, 289 

\item Garijo, A., Athanassoula, E., \& Garcia--Gomez, C. 1997, A\&A, 327, 930

\item Gebhardt, K., et al., 2000, AJ, in press (astro--ph/9912026)

\item Gerhard, O. E., Binney, J. J. 1985, MNRAS, 216, 467

\item Gerhard, O. E. 1993, MNRAS, 265, 213

\item Gerhard, O. E., Jeske, G., Saglia, R. P., Bender, R. 1998, MNRAS, 295,
 197

\item Kochanek, C. S. 1994, ApJ, 436, 56

\item Lawson, C.~L., \& Hanson, R.~J. 1974, Solving Least Squares 
                  Problems (Englewood Cliffs, New Jersey: Prentice--Hall)

\item Matthias, M., \& Gerhard, O. E. 1999, MNRAS, 310, 879

\item Merritt, D. 1993, ApJ, 413, 79

\item Merritt, D., \& Oh, S. P. 1997, AJ, 113, 1279

\item Merritt, D., \& Quinlan, G. D. 1998, ApJ , 498, 625

\item Merritt, D. 1999, Comments on Modern Physics, Vol. 1, 39

\item Navarro, J., Frenk, C., \& White, S. D. M. 1996, ApJ, 462, 563

\item Neistein, E., Maoz, D., Rix, H.-W., Tonry, J. L. 1999, 
AJ, 117, 2666

\item Prugniel, Ph., Heraudeau, Ph. 1998, AAS, 128, 299

\item Press, W.~H., Teukolsky, S.~A., Vetterling, W.~T., \&
        Flannery, B.~P. 1992, Numerical Recipes
        (Cambridge: Cambridge University Press)

\item Richstone, D.~O, et al. 1997,
                  in The Nature of Elliptical Galaxies,
                  Proceedings of the Second Stromlo Symposium,
                  eds. Arnaboldi,~M., da Costa, G., \& Saha, P., p.\ 123

\item Rix, H.--W., \& White, S.~D.~M. 1992, MNRAS, 254, 389
   
\item Rix, H.--W., Kennicutt, R. C., Braun, R., Walterbos, R. A. M. 
		  1995, ApJ, 438, 155

\item Rix, H.-W., de Zeeuw, P. T., Cretton, N., van der Marel, R. P., \&  
		  Carollo, C. M. 1997, ApJ, 488, 702

\item Saglia, R. P., Kronawitter, A., Gerhard, O. E., Bender, R. 1999,
MNRAS, submitted, (astro--ph/9909446)

\item Statler, T. S., Smecker--Hane, T., Cecil, G. N. 1996, AJ, 111, 1512

\item Statler, T. S., Dejonghe, H., Smecker--Hane, T. 1999, AJ, 117, 126 

\item van den Bosch, F. C., \& Emsellem E. 1998, MNRAS, 298, 267

\item van der Marel, R. P. 1991, MNRAS, 253, 710

\item van der Marel, R. P., \& Franx, M. 1993, ApJ, 407, 525

\item van der Marel, R.~P., Cretton, N., de Zeeuw, P.~T., \& Rix, H.~W. 1998,
                  ApJ, 493, 613 

\item Weil, M., Hernquist, L. 1996, ApJ, 460, 101

\end{description}

% ---  Figure 1 ------------------------------------------------------

% ---  Figure 2 ------------------------------------------------------

% ---  Figure 3 ------------------------------------------------------

% ---  Figure 4 ------------------------------------------------------
 
% ---  Figure 5 ------------------------------------------------------
 
% ---  Figure 6  ------------------------------------------------------

% ---  Figure 7 ------------------------------------------------------

% ---  Figure 8 ------------------------------------------------------

% ---  Figure 9 ------------------------------------------------------

% ---  Figure 10 ------------------------------------------------------

% ---  Figure 11 ------------------------------------------------------

% ---  Figure 12 ------------------------------------------------------

% ---  Figure 1 appendix ------------------------------------------------------

%------------------------------------------------------------------------

\newpage

\begin{deluxetable}{cccccccc}
 \tablecaption{Summary of the kinematic observations.\label{t:tbl2320-1}}
\tablehead{
\colhead{$j$} & 
\colhead{name} & 
\colhead{telescope} & 
\colhead{slit width} & 
\colhead{date} & 
\colhead{radial extent} & 
\colhead{\# of points} & 
\colhead{S/N} \\
\colhead{(1)} &
\colhead{(2)} &
\colhead{(3)} &
\colhead{(4)} &
\colhead{(5)} &
\colhead{(6)} &
\colhead{(7)} &
\colhead{(8)} \\
}
\startdata
 1 & MMT-140  &  MMT     & 3.5 & 17 Feb 1996 & 37 & 23 & 25 \\
 2 & MMT-50   &  MMT     & 3.5 & 17 Feb 1996 & 23 & 12 & 25 \\
 3 & 050      &  KPNO 4m & 2.5 & 2  Mar 1998 & 17 & 8  & 12 \\
 4 & 100      &  KPNO 4m & 2.5 & 27 Feb 1998 & 18 & 10 & 10 \\
 5 & 130      &  KPNO 4m & 2.5 & 27 Feb 1998 & 27 & 11 & 12 \\
 6 & 142      &  KPNO 4m & 2.5 & 2  Mar 1998 & 24 & 12 & 12 \\
 7 & 175      &  KPNO 4m & 2.5 & 27 Feb 1998 & 19 & 10 & 10 \\
 8 & b000     &  KPNO 4m & 2.5 & 1  Mar 1998 & 24 & 9  & 15 \\
 9 & b020     &  KPNO 4m & 2.5 & 1  Mar 1998 & 27 & 10 & 15 \\
10 & b100     &  KPNO 4m & 2.5 & 27 Feb 1998 & 45 & 18 & 10 \\
11 & b130     &  KPNO 4m & 2.5 & 27 Feb 1998 & 44 & 19 & 10 \\
12 & b175     &  KPNO 4m & 2.5 & 27 Feb 1998 & 43 & 16 & 10 \\
\enddata
\tablecomments{Column~(1) gives the number of the spectrum;
  column~(2) its name (chosen after its position angle); 
  column~(3) the telescope where it was taken; 
  column~(4) the slit width (in arcseconds); 
  column~(5) the date of observation; 
  column~(6) the maximum radial extension (in arcseconds); 
  column~(7) the number of data points (after pixel binning) for the stellar data; 
  column~(8) the S/N for the stellar data.}
\end{deluxetable} 

% ------------------------------------------------------------------------

% --- Table 2

\begin{deluxetable}{cccc}
\tablecaption{Parameters of MGE model for the luminous density 
 profile.\label{t:tbl2320-2}}
\tablehead{
\colhead{$j$} & 
\colhead{$I_j$} & 
\colhead{$a_j$} & 
\colhead{$q_j$} \\
\colhead{(1)} & 
\colhead{(2)} & 
\colhead{(3)} & 
\colhead{(4)} \\
}
\startdata
 1 & 287764585.49  &  0.820 & 0.544 \\
 2 & 18868477.1    &  3.000 & 0.443 \\
 3 & 1826940.9     &  8.585 & 0.419 \\
 4 & 219268.2      & 19.061 & 0.505 \\
 5 & 9359.8        & 56.370 & 0.676 \\
\enddata
\tablecomments{
 Column~(1) gives the index number of each Gaussian.
  Column~(2) gives its central luminosity density (in $\Lsun/{\rm arcsec}^3$); 
  column~(3) its standard deviation (which expresses the size of the Gaussian along
  the major axis) and column~(4) its flattening.
  All Gaussians have the same position angle and the same center.}
\end{deluxetable}

\end{document}